\documentclass[twocolumn]{autart}

\usepackage{tikz}
\usepackage{amsmath}
\usepackage{latexsym, amssymb}
\usepackage{graphicx}
\usepackage{amsmath, amsbsy}
\usepackage{amsopn, amstext}
\usepackage{hyperref}
\usepackage{cancel, color}
\usepackage{epstopdf}

\DeclareMathOperator{\const}{const.}
\DeclareMathOperator{\Span}{Span}
\DeclareMathOperator{\Col}{Col}
\DeclareMathOperator{\Row}{Row}

\DeclareMathOperator{\lcm}{lcm}

\def\cal{\mathcal}
\def\T{{\bf T}}

\def\diag{diag}
\def\PR{PR}
\def\const{const}

\def\lra{\leftrightarrow}

\def\d{\delta}

\def\D{\Delta}

\def\0{{\bf 0}}
\def\J{{\bf 1}}
\def\T{\mathfrak{T}}

\newcommand{\R}{{\mathbb R}}

\newcommand{\Z}{{\mathbb Z}}

\def\dsum{\mathop{\sum}\limits}
\def\thefootnote{*}

\newtheorem{dfn}[thm]{Definition}
\newtheorem{prp}[thm]{Proposition}
\newtheorem{exa}[thm]{Example}

\begin{document}

\begin{frontmatter}

\title{Invariant and Dual Invariant Subspaces of {\it k}-valued Networks\thanksref{footnoteinfo}} 

\thanks[footnoteinfo]{This work is supported partly by the National Natural Science Foundation of China (NSFC) under Grants   62073315 and 61733018, and the China Postdoctoral Science Foundation 2021M703423 and 2022T150686.}

\author[1,2]{Daizhan Cheng}\ead{dcheng@iss.ac.cn},    
\author[1]{Hongsheng Qi}\ead{qihongsh@amss.ac.cn},               
\author[1,3]{Xiao Zhang}\ead{xiaozhang@amss.ac.cn},
\author[1]{Zhengping Ji}\ead{zpji@amss.ac.cn} 
\address[1]{Key Laboratory of Systems and Control, Academy of Mathematics and Systems Science, Chinese Academy of Sciences}             
\address[2]{Research Center of Semi-tensor Product of Matrices: Theory and Applications, Liaocheng 252026, P.R. China}
\address[3]{National Center for Mathematics and Interdisciplinary Sciences, Academy of Mathematics and Systems Science, Chinese Academy of Sciences} 


\begin{keyword}                           
$k$-valued (control) network; state (or dual) invariant subspace; bearing space; finite ring; finite lattice.               
\end{keyword}                             

\begin{abstract}                          
Consider a $k$-valued network. Two kinds of (control) invariant subspaces, called state and dual invariant subspaces, are proposed, which are subspaces of state space and dual space respectively. Algorithms are presented to verify whether a dual subspace is a dual or dual control invariant subspace. The bearing space of $k$-valued (control) networks is introduced. Using the structure of bearing space, the universal invariant subspace is introduced, which is independent of the dynamics of particular networks. Finally, the relationship between state invariant subspace and dual invariant subspace of a network is investigated. A duality property shows that if a dual subspace is invariant then its perpendicular state subspace is also invariant and vice versa.
\end{abstract}

\end{frontmatter}

\section{Introduction}
Since the state-space approach proposed by Kalman in 1960 \cite{kal60}, the (control) invariant subspace became one of the most important concepts in modern control theory. Consider a linear control system
\begin{align}\label{0.1}
\dot{x}=Ax+Bu,\quad x\in \mathbb{R}^n,u\in \mathbb{R}^m.
\end{align}
A subspace $V\subset \mathbb{R}^n$ is called an $(A,B)$- invariant (or control-invariant) subspace, if $AV\subset V+{\cal B}$, where ${\cal B}$ is a subspace spanned  by $\Col(B)$\cite{won70}. It is clear that this concept is a generalization of $A$ invariant subspace from linear algebra when there is no control. That is,  a subspace $V$ is $A$-invariant if $AV\subset V$.

This concept has then been extended to non-linear control systems. Consider an affine non-linear control system
\begin{align}\label{0.2}
\dot{x}=f(x)+\dsum_{i=1}^mg_i(x)u_i,\quad x\in M,u\in \mathbb{R}^m,
\end{align}
where $M$ is an $n$-dimensional manifold.
Then a distribution ${\cal D}$ is called an $(f,g)$-invariant distribution, if $L_f{\cal D}\subset {\cal D}+{\cal G}$, where $L_f{\cal D}=\{L_f(X)\;|\;X\in {\cal D}\}$ is the set of Lie derivatives of $X$ with respect to $f$, ${\cal G}$ is the distribution generated by $\{g_i\;|\;i=1,\cdots,m\}$ \cite{isi95}. It is also clear that without control, ${\cal D}$ becomes $f$-invariant.

It is well known that $V$ is the control-invariant subspace of (\ref{0.1}), if and only if, there exists a state-feedback control $U=Fx$, such that $V$ is $A+BF$ invariant \cite{won70}. Similar result is also true for (\ref{0.2}) \cite{isi95}. Roughly speaking, a control-invariant subspace is an invariant subspace for a controlled (or closed-loop) system. Hence, the invariant subspace and control-invariant subspace are essentially the same and can be discussed simultaneously.

In general, the invariant subspace plays a fundamental role in system analysis and control design. In most cases, an invariant subspace is established to represent specific properties of a dynamic system. Hence, it is used for solving various control problems, such as decomposition and disturbance decoupling problems; stability and stabilization problems; output tracking problems, etc.

The Boolean network (BN) was firstly proposed by Kauffman to formulate and analyze the genetic regulation networks \cite{kau69}. Many biological systems have exogenous inputs. To deal with them or to manipulate BNs, the Boolean control networks (BCNs) have then been investigated \cite{ide01,dat03}.
In the light of the semi-tensor product (STP) of matrices, the BNs and BCNs have been formulated into the framework of the linear and bi-linear state space model, called the algebraic state space representation (ASSR) \cite{che11,che12}. The last decade or more has witnessed unprecedented progress in the investigation of BNs and BCNs, stimulated by STP.

When the state space approach was introduced to formulate BNs and BCNs, the (control) invariant subspace becomes a useful tool for analyzing the structure and properties of BNs and the control design of BCNs.  Under the regularity assumption, the invariant subspace has been introduced to BNs to solve the disturbance decoupling problem \cite{che11b}. Then a new concept about invariant subspace, which removed the assumption of regularity has been proposed and used to solve decomposition problem \cite{zou15} and disturbance decoupling problem \cite{zhu16}. State feedback invariant subspace has also been proposed and used for designing controls for BCNs \cite{guo15,liu20}.

A recent work \cite{che22} investigates the realization of BNs using their invariant subspace. It reveals that for a BN with $n$ nodes,
since the state variables $X_i(t)\in {\cal D}:=\{0,1\}$, $i\in [1,n]$, the state space should be ${\cal X}={\cal D}^n$.
The set of logical functions,
denoted by ${\cal X}^*:={\cal F}_{\ell}(X_1,X_2,\cdots,X_n)$, is then called the dual space of ${\cal X}$.
It follows that $|{\cal X}^*|=2^{2^n}$, hence the dual space is much larger than the original state space.

Based on this observation, it is easy to find that all the (control) invariant subspaces considered in the literature (precisely speaking, in previously mentioned papers) are subspaces of dual spaces. Hereafter, they should be correctly called the dual (control) invariant subspaces.

A Boolean network approach is the approximation of the active levels of nodes (cells) by quantizing them to $\{0,1\}$. To improve the accuracy of approximation, $k$-valued network is a natural solution \cite{ada03,aku07,jon12}. Using the STP, the ASSR approach for BNs can easily be extended to multi-valued networks \cite{li10}.

The multi-valued network approach is not only useful for biological networks but also very useful for other finite-valued problems such as finite games \cite{guo13,che15}. To see more applications of the multi-valued ASSR approach to finite (networked) games, you are referred to the survey paper \cite{che21}. In fact, the multi-valued ASSR approach can be used for many other finite-value-based problems, such as intelligent problems \cite{zha18}, circuit design \cite{bry84}, etc. Hence, it is worth paying more attention to $k$-valued networks.

When a dynamic system is considered, the state variables take values from a preassigned set. We call it the bearing space. For instance, when linear control system (\ref{0.1}) or nonlinear control system (\ref{0.2}) are considered, the bearing space is $\mathbb{R}$. When a BN is considered, the bearing space is the Boolean set ${\cal D}:=\{0,1\}$.

The structures and properties of the bearing space do affect the properties of a dynamic system over the space. For BNs, Boolean algebra has been developed to describe the bearing space of BNs \cite{ros03}. Similarly, a $k$-valued algebra, proposed firstly by Post and then named by Post algebra, has been developed to describe the bearing space of $k$-valued networks \cite{pos21,eps60,rem06,ric07,tho01}.

Post lattice (or Post algebra) is not the only structure for bearing space of $k$-valued networks.
Recently, when the multi-valued networks are deeply investigated, people started to pay more attention to the structure of bearing spaces of the networks, and various bearing spaces are proposed.  For instance, the networks over finite fields have been studied by several authors \cite{sun12,pas14,li16,li18,li19,men20}. It follows that the networks over finite rings and networks over finite lattices have also been investigated \cite{che22,jipr}.

 Exploring some bearing spaces of $k$-valued networks, we found some state invariant subspaces appear naturally and these state invariant subspaces play a fundamental role in both structure analysis and control design for $k$-valued networks. This is the motivation for us to investigate both dual and state (control) invariant subspaces of $k$-valued (control) networks.

The main work and contribution of this paper consist of the following:
\begin{itemize}
\item[(i)] Though the concept of  (control) invariant subspaces has been used for a considerable long time, they have been classified clearly into two kinds of invariant subspaces for the first time, which clarified some confusing points in the past use. 
\item[(ii)] An algorithm is presented to verify whether a dual subspace of a $k$-valued network is a dual control-invariant subspace. If the answer is ``yes", the algorithm can also produce the state-feedback controls to make the subspace a dual invariant subspace for the closed-loop network. The algorithm can also be used for BCNs. To our best knowledge, there was no known algorithm for BCNs.
\item[(iii)] By investigating bearing space and its bearing subspace, the concept of universal state invariant subspace is proposed. Correspondingly, a universal cascading normal form is obtained. Here ``universal" means that the invariant subspace and the normal form are independent of the dynamics of individual networks, while they depend on the structure of bearing spaces only.
\item[(iv)] The duality relationship of state subspaces with their perpendicular dual subspaces is revealed, which shows that if a state subspace is invariant then its perpendicular dual subspace is also invariant, and vice versa. This relationship leads to the discovery of universal dual invariant subspaces.
\end{itemize}

The rest of this paper is outlined as follows: Section 2 investigates dual (control) invariant subspace. Necessary and sufficient conditions are obtained for dual control invariant subspaces. Two algorithms are proposed to numerically verify whether a subspace is a dual (or a dual control) invariant subspace. As a special kind of dual (control) subspaces, the node invariant subspaces are also studied. Section 3 considers the state invariant subspaces, including the attractor-based state invariant subspaces and the bearing space-based invariant subspaces.
 The universal state invariant subspace is investigated in Section 4. Using universal state invariant subspace the universal cascading normal form is obtained. Section 5 reveals the duality of state invariant subspace with dual invariant subspace. Using duality, the universal dual invariant subspace is also discovered. Section 6 is a brief summary. Finally, an appendix is attached, which consists of two parts: Part 1 is a quick survey on the STP of matrices and the ASSR of networks. Part 2 is a brief introduction to some bearing spaces, including field, ring, and lattice.

Before ending this section, we give a list of notations:

\begin{enumerate}

\item  ${\cal M}_{m\times n}$: the set of $m\times n$ real matrices.

\item $\Col(M)$ ($\Row(M)$): the set of columns (rows) of $M$. $\Col_i(M)$ ($\Row_i(M)$): the $i$-th column (row) of $M$.

\item ${\cal D}_k:=\left\{1,2,\cdots,k\right\},\quad k\geq 2$.

\item $\d_n^i$: the $i$-th column of the identity matrix $I_n$.

\item $\D_n:=\left\{\d_n^i\vert i=1,\cdots,n\right\}$.

\item ${\bf 1}_{\ell}=(\underbrace{1,1,\cdots,1}_{\ell})^\mathrm{T}$.

\item ${\bf 0}_{p\times q}$: a $p\times q$ matrix with zero entries.

\item A matrix $L\in {\cal M}_{m\times n}$ is called a logical matrix
if the columns of $L$ are of the form of
$\d_m^k$. That is, $
\Col(L)\subset \D_m$.
Denote by ${\cal L}_{m\times n}$ the set of $m\times n$ logical
matrices.

\item $L=[\d_n^{i_1},\d_n^{i_2},\cdots,\d_n^{i_r}]\in {\cal L}_{n\times r}$  is briefly denoted as $
L=\d_n[i_1,i_2,\cdots,i_r]$.

\item $\ltimes$: the semi-tensor product of matrices.

\item $*$: Khatri-Rao product of matrices.

\item ${\PR}_k:=\diag(\d_k^1,\d_k^2,\cdots,\d_k^k)$ is called the $k$-th  power-reducing matrix.

\item $W_{[m,n]}:=[I_n\otimes \d_m^1, I_n\otimes \d_m^2,\cdots,I_n\otimes \d_m^m]$ is called the $(m,n)$-th swap matrix.

\item $\wedge,\vee,\neg$: The conjunction, disjunction, negation operators. \\

\end{enumerate}

\section{Two Kinds of Invariant Subspaces}

\subsection{Invariant Dual Subspace}

Consider a $k$-valued network
\begin{align}\label{2.1.1}
\begin{cases}
X_1(t+1)=f_1(X_1(t),\cdots,X_n(t)),\\
X_2(t+1)=f_2(X_1(t),\cdots,X_n(t)),\\
\vdots\\
X_n(t+1)=f_n(X_1(t),\cdots,X_n(t)),\\
\end{cases}
\end{align}
where $X_i\in {\cal D}_k$, $f_i:{\cal D}_k^n\rightarrow {\cal D}_k$ are $k$-valued logical functions, $i\in [1,n]$.\\
Its control version is
\begin{align}\label{2.1.2}
\begin{cases}
X_1(t+1)=f_1(X_1(t),\cdots,X_n(t),U_1(t),\cdots,U_m(t)),\\
X_2(t+1)=f_2(X_1(t),\cdots,X_n(t),U_1(t),\cdots,U_m(t)),\\
\vdots\\
X_n(t+1)=f_n(X_1(t),\cdots,X_n(t),U_1(t),\cdots,U_m(t)),\\
\end{cases}
\end{align}
where $U_j\in {\cal D}_k,~j=1,2,\cdots,m$ are controls.

Assume $X_i=j\in[1,k]$, its vector form expression is denoted by
$$
x_i=\vec{X}_i:=\d_k^j.
$$
Using vector form expression, $k$-valued network (\ref{2.1.1}) can be expressed by its component-wise ASSR as
\begin{align}\label{2.1.3}
x_i(t+1)=M_ix(t),\quad i\in [1,n],
\end{align}
where $x(t)=\ltimes_{j=1}^nx_j(t)$, $M_i\in {\cal L}_{k\times k^n}$ is the structure matrix of $f_i$. Moreover, the overall ASSR of (\ref{2.1.1}) is
\begin{align}\label{2.1.4}
x(t+1)=Mx(t),
\end{align}
where $M=M_1*M_2*\cdots*M_n\in {\cal L}_{k^n\times k^n}$.

Similarly, the $k$-valued control network (\ref{2.1.2}) can be expressed by its component-wise ASSR as
\begin{align}\label{2.1.5}
x_i(t+1)=L_iu(t)x(t),\quad i\in [1,n],
\end{align}
where $u(t)=\ltimes_{j=1}^mu_j(t)$, $L_i\in {\cal L}_{k\times k^{n+m}}$ is the structure matrix of $f_i$. Moreover, the overall ASSR of (\ref{2.1.2}) is
\begin{align}\label{2.1.6}
x(t+1)=Lu(t)x(t),
\end{align}
where $L=L_1*L_2*\cdots*L_n\in {\cal L}_{k^n\times k^{n+m}}$.

The invariant subspace of network (\ref{2.1.1}) or control networks (\ref{2.1.2}) is generated by some $k$-valued logical functions. In a recent work \cite{che22}, the concept of invariant subspace for Boolean (control) networks has been clearly stated and certain fundamental properties have also been provided. This concept and properties can be easily extended to $k$-valued networks. So we just state them in the $k$-valued version.

Denote by ${\cal X}=\{X_1,X_2,\cdots,X_n\}$ the state space of the networks; ${\cal F}_{\ell}(X_1,X_2,\cdots,X_n)$ the set of $k$-valued logical functions with arguments $X_1,X_2,\cdots,X_n$; ${\cal X}^*:={\cal F}_{\ell}(X_1,X_2,\cdots,X_n)$ the dual space of ${\cal X}$. Then we have the following definition.

\begin{dfn}\label{d2.1.1} Consider the $k$-valued network (\ref{2.1.1}). Let $\{Z_1,Z_2,\cdots,Z_r\}\subset {\cal X}^*$ be a set of logical functions. Define ${\cal Z}:={\cal F}_{\ell}(Z_1,Z_2,\cdots,Z_r)$ the subspace generated by  $\{Z_1,Z_2,\cdots,Z_r\}$. ${\cal Z}$ is said to be an invariant (dual) subspace of (\ref{2.1.1}) if $Z(0)=(Z_1(0),Z_2(0),\cdots,Z_r(0))\in {\cal Z}$ implies $Z(t)\in {\cal Z},~\forall t\in\mathbb{Z}_+$.
\end{dfn}

Let $Z_i=g_i(X_1,X_2,\cdots,X_n)$ with $z_i=\vec{Z}_i$ and the structure matrix of $g_i$ be $G_i\in {\cal L}_{k\times k^n}$. Then
\begin{align}\label{2.1.7}
z:=\ltimes_{i=1}^rz_i=Gx,
\end{align}
where $G=G_1*G_2*\cdots*G_r\in {\cal L}_{k^r\times k^n}$.

Now we have the following conclusion, which is an immediate extension of the corresponding result in \cite{che22}:

\begin{prp}\cite{che22}\label{p2.1.2} ${\cal Z}$ is an invariant dual subspace of (\ref{2.1.1}), if and only if, there exists a logical matrix $H\in {\cal L}_{k^r\times k^r}$ such that
\begin{align}\label{2.1.8}
GM=HG.
\end{align}
Moreover, let $z(t)=Gx(t)$. Then the dynamics of $z(t)$ is determined by
\begin{align}\label{2.1.9}
z(t+1)=Hz(t).
\end{align}
\end{prp}

${\cal Z}$ is also called the $M$-invariant dual subspace.

\begin{dfn}\label{d2.1.3} Consider the $k$-valued control network (\ref{2.1.2}). Let $\{Z_1,Z_2,\cdots,Z_r\}\subset {\cal X}^*$ be a set of logical functions. Define ${\cal Z}:={\cal F}_{\ell}(Z_1,Z_2,\cdots,Z_r)$ the subspace generated by  $\{Z_1,Z_2,\cdots,Z_r\}$. ${\cal Z}$ is said to be control-invariant dual subspace of (\ref{2.1.2}) if for any $X(0)=(X_1(0),X_2(0),\cdots,X_n(0))\in {\cal Z}$ there exists a control sequence $\{U(t)\;|\;t\geq 0\}$  such that ${\cal Z}$ is invariant to the controlled network. In other words, the controlled network satisfies $X(t)\in {\cal Z}$.
\end{dfn}

Splitting $L$ into $k^m$ equal blocks as
$$
L=[M_1,M_2,\cdots,M_{k^m}],
$$
where
$
M_i=L\d_{k^m}^i\in {\cal L}_{k^n\times k^n},~i\in[1,k^m].
$

For a fixed $H\in {\cal L}_{k^r\times k^r}$, we define a set of index sets as follows:
\begin{align}\label{2.1.10}
J_j:=\left\{i\;|\;\Col_j(GM_i)=\Col_j(HG)\right\},\quad j\in [1,k^n].
\end{align}

\begin{thm}\label{t2.1.4}  Consider the $k$-valued control network (\ref{2.1.2}). ${\cal Z}:={\cal F}_{\ell}(Z_1,Z_2,\cdots,Z_r)$ is  a dual control-invariant subspace of (\ref{2.1.2}), if and only if, one of the following two equivalent conditions is satisfied.
\begin{itemize}
\item[(i)] There exists an $H\in {\cal L}_{k^r\times k^r}$, such that $\forall j\in [1,k^n]$,
\begin{align}\label{2.1.11}
J_j\neq \emptyset.
\end{align}
\item[(ii)] There exists a state feedback control
\begin{align}\label{2.1.12}
u(t)=Fx(t),
\end{align}
where $F\in {\cal L}_{k^m\times k^n}$, such that
${\cal Z}$ is $M$-invariant, where $M:=L F {\PR}_{k^n}$.
\end{itemize}
 \end{thm}

{\it Proof:}
(Necessity) According to Proposition \ref{p2.1.2}, at each time $t$ there must exist a $u(t)$ and an $H\in  {\cal L}_{k^r\times k^r}$, such that
for the controlled system,
\begin{align}\label{2.1.13}
z(t+1)=Hz(t).
\end{align}
Assume $u(t)=\d_{k^m}^i$, then (\ref{2.1.13}) leads to
\begin{align}\label{2.1.14}
GM_ix(t)=HGx(t).
\end{align}
Assume $x(t)=\d_{k^n}^j$, then (\ref{2.1.14}) implies $i\in J_j$, that is, $J_j\neq \emptyset$.

Next, from (\ref{2.1.11}) we know that for each $x(t)=\d_{k^n}^j$ we can find that $u(t)=\d_{k^m}^i$, where $i\in J_j$ may not be unique, such that (\ref{2.1.13}) holds. Then the state-feedback matrix $F$ exists.

(Sufficiency) Assume  (\ref{2.1.11}) holds. We can construct state feedback control as aforementioned. Using this state feedback control, (\ref{2.1.14}) ensures $x(t+1)\in {\cal Z}$ as long as $x(t)\in {\cal Z}$.
 \hfill $\Box$

Next, we look for numerical algorithms to verify whether a dual subspace is (control) invariant.

To begin with, observing condition (\ref{2.1.8}), one can search $H$ as follows: Assume $\Col_j(G)=\d_{k^r}^{\ell_j}$, then
$\Col_j(HG)=\Col_{\ell_j}(H)$. Hence, to satisfy (\ref{2.1.8}), it is enough to choose
\begin{align}\label{2.1.15}
\Col_{\ell_j}(H)=\Col_{j}(GM),\quad j\in[1,k^n].
\end{align}

The following corollary is obvious:

\begin{cor}\label{c2.1.5} Consider the $k$-valued network (\ref{2.1.1}). Define a set of index sets as
\begin{align}\label{2.1.15}
S_i:=\left\{j\;|\; \Col_j(G)=\d_{k^r}^i\right\},\quad i\in [1,k^r].
\end{align}
Then ${\cal Z}$ is $M$-invariant, if and only if,
\begin{align}\label{2.1.16}
\Col_j(GM)=\const.,\quad j\in S_i.
\end{align}
\end{cor}

{\it Proof:} Necessity comes from the fact that (\ref{2.1.15}) forces (\ref{2.1.16}) to be true. As for the sufficiency, as long as (\ref{2.1.16}) holds, we can simply set
\begin{align}\label{2.1.17}
\Col_i(H):=\Col_j(GM),\quad i\in [1,k^r],
\end{align}
which verifies (\ref{2.1.8}).
\hfill $\Box$

\begin{rem}\label{r2.1.501} It $S_i=\emptyset$, Eq. (\ref{2.1.16}) is automatically true. Then the corresponding $\Col_i(H)$ can be chosen arbitrarily.
\end{rem}

The algorithm follows immediately.

\begin{alg}\label{a2.1.6} Let ${\cal Z}$ be a dual subspace of (\ref{2.1.1}) and its structure matrix be $G\in L_{k^r\times k^n}$.

\begin{itemize}

\item Step 1: Construct $S_i$, $i\in [1,k^r]$ by (\ref{2.1.15}).

\item Step 2: Check if (\ref{2.1.16}) holds.

If the answer is ``yes", ${\cal Z}$ is $M$-invariant. ($H$ can be constructed by (\ref{2.1.17}).)

Otherwise, ${\cal Z}$ is not $M$-invariant.

\end{itemize}
\end{alg}

Consider the control network (\ref{2.1.2}).

Observing condition (\ref{2.1.11}) one can search $H$ as follows: Assume $\Col_j(G)=\d_{k^r}^{\ell_j}$, then
$\Col_j(HG)=\Col_{\ell_j}(H)$. According to (\ref{2.1.14}), there exists at least one $i$ such that  (\ref{2.1.15}) holds for $M=M_i$. It follows that
\begin{align}\label{2.1.18}
\Col_{\ell_j}(H)\in \bigcup_{i=1}^{k^m}\Col_{j}(GM_i),\quad j\in[1,k^n].
\end{align}

Hence, we have the following conclusion.

\begin{cor}\label{c2.1.7} Consider the control network (\ref{2.1.2}). ${\cal Z}$ is control invariant, if and only if,
\begin{align}\label{2.1.19}
\bigcap_{j\in S_i}\left(\bigcup_{\ell=1}^{k^m}\Col_j(GM_{\ell})\right)\neq \emptyset,\quad i\in [1,k^r].
\end{align}
\end{cor}

Then the corresponding algorithm can be obtained as follows:

\begin{alg}\label{a2.1.8} Let ${\cal Z}$ be a dual subspace of (\ref{2.1.2}) and its structure matrix be $G\in L_{k^r\times k^n}$.

\begin{itemize}

\item Step 1: Construct $S_i$, $i\in [1,k^r]$ by (\ref{2.1.15}).

\item Step 2: Check if (\ref{2.1.19}) holds.

If the answer is ``yes", ${\cal Z}$ is the control invariant. Go to Step 3.

Otherwise, ${\cal Z}$ is not control invariant. Stop.

\item Step 3: (Construct state feed control) For each $i\in[1,k^r]$, choose a (anyone)
$$
\xi_i\in \bigcap_{j\in S_i}\left(\bigcup_{\ell=1}^{k^m}\Col_j(GM_{\ell})\right)
$$
Then for each $j$, there is (at least) a $\ell_j\in [1,k^m]$, such that
$$
\xi_i=\Col_j(GM_{\ell_j}).
$$
Set
\begin{align}\label{2.1.20}
u=\d_{k^m}^{\ell_j},\quad x=\d_{k^n}^j,\; j\in S_i.
\end{align}

\end{itemize}
\end{alg}

Note that Remark \ref{r2.1.501} remain true. That is, if $S_i=\emptyset$, then $\xi_i$ can be arbitrary.

\begin{rem}\label{r2.1.9} Using (\ref{2.1.20}), the state feedback matrix $F$ as in (\ref{2.1.12}) is obtained. By construction, it is ready to verify that ${\cal Z}$ is $M=LF{\PR}_{k^n}$ invariant. That is, control invariant.
\end{rem}

\begin{exa}\label{e.2.1.10}
Consider the $3$-valued control network
\begin{align}\label{2.1.21}
\begin{cases}
X_1(t+1)=X_2(t)\vee U(t),\\
X_2(t+1)=X_1(t)\wedge U(t).
\end{cases}
\end{align}

Assume ${\cal Z}={\cal F}_{\ell}\{Z\}$, where
$$
z=\vec{Z}=\d_3[1,2,1,1,3,3,2,3,2]x:=Gx.
$$
Check whether ${\cal Z}$ is a control dual invariant subspace.

The structure matrices of $\vee$ and $\wedge$ are as follows:
$$
\begin{array}{l}
M_d:=M_{\vee}=\d_3[1,1,1,1,2,2,1,2,3],\\
M_c:=M_{\wedge}=\d_3[1,2,3,2,2,3,3,3,3].
\end{array}
$$
Then the component-wise ASSR is
\begin{align}\label{2.1.22}
\left\{\begin{array}{ccl}
x_1(t+1)&=&M_du(t)x_2(t)\\
~&=&M_d(I_3\otimes {\bf 1}_3^\mathrm{T}\otimes I_3)u(t)x(t)\\
~&:=&L_1u(t)x(t),\\
x_2(t+1)&=&M_cu(t)x_1(t)\\
~&=&M_c(I_9\otimes {\bf 1}_3^\mathrm{T})u(t)x(t)\\
~&:=&L_2u(t)x(t),
\end{array}
\right.
\end{align}
where
$$
\begin{array}{ccl}
L_1&=&\d_3[1,1,1,1,1,1,1,1,1,1,2,2,1,2,\\
~&~&~~~2,1,2,2,1,2,3,1,2,3,1,2,3].\\
L_2&=&\d_3[1,1,1,2,2,2,3,3,3,2,2,2,2,2,\\
~&~&~~~2,3,3,3,3,3,3,3,3,3,3,3,3].\\
\end{array}
$$
Then
$$
\begin{array}{ccl}
L&=&L_1*L_2\\
~&=&\d_9[1,1,1,2,2,2,3,3,3,2,5,5,2,5,\\
~&~&~~~5,3,6,6,3,6,9,3,6,9,3,6,9].\\
\end{array}
$$

Next, we calculate $GM_i$, where $M_i=L\d_3^i$, $i=1,2,3$. Put them into Table \ref{Tab2.1.1}.

\vskip 5mm

\begin{table}[!htbp]
  \centering
  \caption{Searching Common Vectors in $GM_i$\label{Tab2.1.1}}
  \doublerulesep 0.5pt
  \begin{tabular}{|c||c|c|c|c|c|c|c|c|c|}
    \hline $j$ &$1$&$2$&$3$&$4$&$5$&$6$&$7$&$8$&$9$\\
    \hline $\Col_j(GM_1)$ & $\delta_3^1$ & $\underline{\delta_3^1}$ & $\delta_3^1$ & $\delta_3^2$  & $\delta_3^2$ & $\delta_3^2$& $\delta_3^1$ & $\delta_3^1$ & $\underline{\delta_3^1}$ \\
    \hline $\Col_j(GM_2)$ & $\underline{\delta_3^2}$ & $\delta_3^3$ & $\delta_3^3$ & $\underline{\delta_3^2}$  & $\delta_3^3$ & $\underline{\delta_3^3}$& $\delta_3^1$ & $\delta_3^3$ & $\delta_3^3$\\
     \hline $\Col_j(GM_3)$ & $\delta_3^1$ & $\delta_3^3$ & $\underline{\delta_3^2}$ & $\delta_3^1$  & $\underline{\delta_3^3}$ & $\delta_3^2$& $\underline{\delta_3^1}$ & $\underline{\delta_3^3}$ & $\delta_3^2$\\
    \hline
  \end{tabular}
\end{table}

\vskip 5mm

Now we search for possible $H=[h_1,h_2,h_3]$:

Since
$$
G=\d_3[1,2,1,1,3,3,2,3,2],
$$
we have
$$
\begin{array}{l}
S_1=\{1,3,4\},\\
S_2=\{2,7,9\},\\
S_3=\{5,6,8\}.
\end{array}
$$
Consider $S_1$, we have
$$
h_1\in \bigcap_{j=1,3,4}\{\Col_j(GM_1)\bigcup \Col_j(GM_2) \bigcup \Col_j(GM_3)\}.
$$
Then we can have $h_1\in \{\d_3^1,\d_3^2\}$. We choose, say, $h_1=\d_3^2$. Correspondingly, for $j=1$ we can choose $M_2$,  for $j=3$ we can choose $M_3$,
for $j=4$ we can choose either $M_1$ or $M_2$, say, we choose $M_2$. Then we underline the corresponding $M_i$ in Table \ref{Tab2.1.1}.

Similarly, considering $S_2$, we have
$$
h_2\in \bigcap_{j=2,7,9}\{\Col_j(GM_1)\bigcup \Col_j(GM_2) \bigcup \Col_j(GM_3)\}.
$$
Then we have $h_2=\d_3^1$. Corresponding $M_i$ are also underlined in the Table.

Finally, for $S_3$ we have
$$
h_3\in \bigcap_{j=5,6,8}\{\Col_i(GM_1)\bigcup \Col_j(GM_2) \bigcup \Col_j(GM_3)\}.
$$
Then we have $h_3=\d_3^3$. Corresponding $M_i$ are also underlined in the Table.

According to Theorem \ref{t2.1.4}, we conclude that ${\cal Z}$ is a control-invariant dual subspace. Moreover, using (\ref{2.1.20}), the state feedback can be constructed as
$$
u(x)=
\begin{cases}
\d_3^1,\quad x=\d_9^2,\d_9^9,\\
\d_3^2,\quad x=\d_9^1,\d_9^4,\d_9^6,\\
\d_3^3,\quad x=\d_9^3,\d_9^5,\d_9^7,\d_9^8.\\
\end{cases}
$$
Equivalently,
$$
u(x)=Fx,
$$
where
$
F=\d_3[2,1,3,2,3,2,3,3,1].
$

A straightforward verification shows that
$$
GLF{\PR}_{9}=HG,
$$
where
$
H=\d_3[2,1,3].
$

From the above process, one sees easily that the state feedback control is not unique.
\end{exa}

\subsection{Node Invariant Subspace}

\begin{dfn}\label{d2.2.1}
\begin{itemize}
\item[(i)] Assume the $k$-valued network (\ref{2.1.1}) can be expressed by
\begin{align}\label{2.2.1}
\begin{cases}
X^1(t+1)=F_1(X^1(t)),\\
X^2(t+1)=F_2(X^1(t),X^2(t)),\\
\end{cases}
\end{align}
where $(X^1,X^2)$ is a partition of $\{X_1,X_2,\cdots,X_n\}$. Then $X^1$ is called a node invariant subspace. \footnote{The node-invariant subspace was called a regular invariant subspace in previous literature, e.g., \cite{che11b}}.

\item[(ii)] Consider the $k$-valued control network (\ref{2.1.2}). Assume under a suitable (state feedback) control
$U(t)=GX(t)$, the closed-loop network of  (\ref{2.1.2}) can be expressed into the form of (\ref{2.2.1}), then $X^1$ is called a node control-invariant subspace.
\end{itemize}
\end{dfn}

Both node invariant subspace and node control-invariant subspace are particularly useful in decoupling problems and has been discussed in detail by \cite{che11b} under the name of regular subspace.

To verify whether a subspace is a node invariant subspace the key issue is to know when a subspace is a node subspace. The following result is fundamental.

\begin{prp}\label{p2.2.2} \cite{che11b} Consider the $k$-valued network (\ref{2.1.1}) with its ASSR (\ref{2.1.4}). Let ${\cal H}^*=\{h_1,h_2,\cdots,h_s\} \subset {\cal X}^*$ be a dual subspace, and $H$ is the structure matrix of ${\cal H}^*$.  ${\cal H}$ is a node (or regular) subspace, if and only if,
\begin{align}\label{2.2.2}
\left|\left\{j\;|\; \Col_j(H)=\d_{k^s}^i\right\}\right|=k^{n-s},\quad \forall i\in [1,k^s].
\end{align}
\end{prp}

\begin{rem}\label{r2.2.2}
\begin{itemize}
\item[(i)] Both node invariant subspace and node control-invariant subspace are special cases of dual invariant subspace and dual control-invariant subspace respectively. The only special point of node invariant subspaces lies in that the subspaces are generated by a set of coordinate functions.

\item[(ii)] To get the decomposed form  (\ref{2.2.1}) a coordinate change may be required.

\item[(iii)] Intuitively, a node invariant subspace ${\cal H}^*$ is a set of nods, which form a subgraph in the network graph. Moreover, the in-degree
of the subgraph is zero. (Refer to Fig. \ref{Fig.2.2.1}).
\begin{figure}
\centering
\setlength{\unitlength}{10mm}
\begin{picture}(8,3)\thicklines
\put(1,1){\circle*{0.2}}
\put(1,2){\circle*{0.2}}
\put(2,1){\circle*{0.2}}
\put(2,2){\circle*{0.2}}
\put(4,1){\circle*{0.2}}
\put(4,2){\circle*{0.2}}
\put(5,1){\circle*{0.2}}
\put(5,2){\circle*{0.2}}
\put(1,1){\vector(1,1){0.9}}
\put(1,1){\vector(1,0){0.9}}
\put(1,1){\vector(0,1){0.9}}
\put(1,2){\vector(1,0){0.9}}
\put(2,1){\vector(0,1){0.9}}
\put(2,1){\vector(1,0){1.9}}
\put(2,2){\vector(1,0){1.9}}
\put(4,1){\vector(1,0){0.9}}
\put(4,2){\vector(1,0){0.9}}
\put(4,2){\vector(1,-1){0.9}}
\put(5,2){\vector(1,0){0.5}}
\put(5,1){\vector(1,0){0.5}}
\put(5.8,0.9){$\cdots$}
\put(5.8,1.9){$\cdots$}
\put(0.6,2.2){${\cal H}^*$}
\thinlines
\put(0.5,0.5){\line(1,0){2}}
\put(0.5,0.5){\line(0,1){2}}
\put(2.5,2.5){\line(-1,0){2}}
\put(2.5,2.5){\line(0,-1){2}}
\put(3.5,0.5){\line(1,0){3}}
\put(3.5,0.5){\line(0,1){2}}
\put(6.5,2.5){\line(-1,0){3}}
\put(6.5,2.5){\line(0,-1){2}}
\end{picture}
\caption{Node Subspace ${\cal H}^*$ \label{Fig.2.2.1}}
\end{figure}
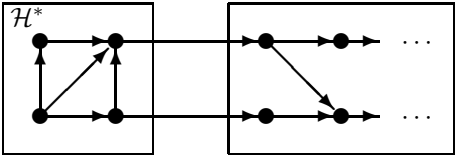

\item[(iv)] For a control network, to verify the node control-invariant subspace the method provided for the dual invariant subspace is also applicable. The only difference lies in verifying whether the subspace is a node (or regular) one.
\end{itemize}
\end{rem}

\section{State Invariant Subspace}

\begin{dfn}\label{d2.3.1}
\begin{itemize}
\item[(i)] Consider the $k$-valued network (\ref{2.1.1}). Assume $V\subset {\cal D}_k^n$ is a subset of the state space ${\cal X}$. Let $X(0)=\{X_1(0),X_2(0),\cdots,X_n(0)\}\in V$. If the trajectory $X(t,X(0))\in V$, $t\geq 0$, then $V$ is called a state invariant subspace of (\ref{2.1.1}).
\item[(ii)] Consider the $k$-valued control network (\ref{2.1.2}). Assume $V\subset {\cal D}_k^n$ be a subset of the state space ${\cal X}$. Let $X(0)=\{X_1(0),X_2(0),\cdots,X_n(0)\}\in V$. Assume there is a control sequence $\{U(t)\;|\; t\geq 0\}$ If the trajectory $X(t,U(0),X(0))\in V$, $t\geq 0$, then $V$ is called a state control-invariant subspace of (\ref{2.1.2}).
\end{itemize}
\end{dfn}

\begin{prp}\label{p2.3.2} Let $V\subset {\cal X}$ with $|V|=r$.
\begin{itemize}
\item[(i)] $V$ is invariant with respect to network (\ref{2.1.1}) (or $M$-invariant), if and only if, there exists a logical matrix $E\in {\cal L}_{r\times r}$ such that
\begin{align}\label{2.3.1}
MV=VE.
\end{align}
\item[(ii)] $V$ is control-invariant to the control network (\ref{2.1.2}), if and only if, there exists a state feedback
\begin{align}\label{2.3.2}
u(t)=Gx(t),
\end{align}
where $G\in {\cal L}_{k^m\times k^n}$, such that $V$ is $LG{\PR}_{k^n}$-invariant.
\end{itemize}
\end{prp}

{\it Proof:}
\begin{itemize}
\item[(i)] The existence of a real matrix $E$ is well known in linear algebra. Since each column of $MV$ is a logical vector, it must be a column of $V$, which ensures $E$ to be a logical matrix.
\item[(ii)] The argument is similar to the proof of Theorem \ref{t2.1.4}.
\end{itemize}
\hfill $\Box$

State invariant subspace is closely related to the topological structure of networks. The normal form of Boolean networks was firstly revealed by \cite{for13}. The following form comes from \cite{for13,liu19}, which is for Boolean networks. It is also true for $k$-valued networks:

\begin{prp}\label{p2.3.3} Consider the $k$-valued network (\ref{2.1.1}). Under a suitable coordinate frame $z=Tx$, its ASSR can be expressed as
\begin{align}\label{2.3.3}
z(t+1)=\tilde{M}z(t),
\end{align}
where
$$
\tilde{M}=\diag\left(\tilde{M}_1,\tilde{M}_2,\cdots,\tilde{M}_k\right),
$$
and each block can be expressed as
$$
\tilde{M}_j=\begin{bmatrix}
C_j&D_j\\
0,N_j
\end{bmatrix},\quad j\in[1,k],
$$
where $C_j$ is a cyclic matrix, that is,
$$
C_j=\d_{n_j}[2,3,\cdots,n,1],
$$
corresponding to a cycle of length $n_j$; $N_j$ is a nilpotent matrix, that is, there exists an $s_j>0$ such that $(N_j)^{s_j}=0$, corresponding to the basin of attraction of $C_j$.
\end{prp}

Since a coordinate transformation of networks is only a state rename, it does not change the topological structure (including attractors and their basins of attraction), the structure of normal form represents general forms. From the normal form, it is clear that a state invariant subspace of a network consists of several attractors with (might be a subset of) their basins of attraction. This is the only form of state invariant subspaces.

State (control) invariant subspaces have been applied to  set reachability \cite{zho19}, set stability \cite{guo17,gou19}, output regulation \cite{zha20}, etc. under terminologies of ``invariant set" and ``control invariant set".  We omit further discussion here.

In the following, we consider a special kind of state invariant subspaces, which are caused by the structure of the bearing space. The first example considers a network over a finite ring \cite{che22b}.

\begin{exa}\label{e2.3.4} Consider a $6$-valued network as
\begin{align}\label{2.3.3}
\begin{cases}
X_1(t+1)=p_1(X_1(t),X_2(t),\cdots,X_n(t)),\\
X_2(t+1)=p_2(X_1(t),X_2(t),\cdots,X_n(t)),\\
\cdots,\\
X_n(t+1)=p_n(X_1(t),X_2(t),\cdots,X_n(t)),\\
\end{cases}
\end{align}
where $X_i(t)\in \Z_6$, $p_i$ are polynomials, $i\in[1,n]$.

Note that $\Z_6=\{0,1,2,3,4,5\}.$ The addition $\oplus$ and the product is $\odot$ and defined as follows: 
\begin{align*}
\begin{array}{l}
a\oplus b:=a+b(mod~6),\\
a\odot b:=ab(mod~6).
\end{array}
\end{align*}

Under these $a\oplus b$ and $a\odot b$, $\Z_6$ is a commutative ring with identity $1$, and the functions over $\Z_6$ can be expressed as polynomials.
For system (\ref{2.3.3}) the state space is
$$
{\cal X}=\{X_1,X_2,\cdots,X_n\}.
$$
In vector form, set
$$
i\sim \begin{cases}
\d_6^i,\quad i\in [1,5],\\
\d_6^6,\quad i=0.
\end{cases}
$$
Then the state space can be expressed as
$$
{\cal X}=\D_6^n.
$$
It is well known that
$$
J:=\{0,3\}\sim\{\d_6^6,\d_6^3\}
$$
is an ideal of $\Z_6$. If each $X_i\in J$, then we have a subset
$$
J^n=\{\d_6^3,\d_6^6\}^n\subset \D_6^n,
$$
which is a state subspace of ${\cal X}$. Moreover, Since $J$ is an ideal of $\Z_6$,  $J^n$ is a state invariant subspace of the network (\ref{2.3.1}).
\end{exa}

The next example is a network over finite lattice \cite{jipr}.

\begin{exa}\label{e2.3.5} Consider a control network over lattice $L$ as
\begin{align}\label{2.3.4}
\begin{array}{l}
\begin{cases}
X_1(t+1)=X_2(t)\vee U(t),\\
X_2(t+1)=X_3(t)\wedge X_1(t),\\
X_3(t+1)=\left(X_1(t)\vee X_2(t)\right)\wedge X_3(t),
\end{cases}\\
~~~Y(t)=X_2(t)\vee X_3(t).
\end{array}
\end{align}
The lattice $L$ has its Hasse diagram as in Fig. \ref{Fig.2.3.1}.

\vskip 5mm

\begin{figure}
\centering
\setlength{\unitlength}{5mm}
\begin{picture}(4,9)(-1,-3)\thicklines
\put(0,-2){\circle*{0.2}}
\put(0,0){\circle*{0.2}}
\put(0,4){\circle*{0.2}}
\put(2,2){\circle*{0.2}}
\put(2,6){\circle*{0.2}}
\put(0,0){\line(1,1){2}}
\put(0,0){\line(0,-1){2}}
\put(0,4){\line(1,1){2}}
\put(0,0){\line(0,1){4}}
\put(2,2){\line(0,1){4}}
\put(-0.8,-2){$\d_5^5$}
\put(-0.8,0){$\d_5^4$}
\put(-0.8,4){$\d_5^2$}
\put(2.2,2){$\d_5^3$}
\put(2.2,6){$\d_5^1$}
\end{picture}
\caption{Hasse Diagram of $L$ \label{Fig.2.3.1}}
\end{figure}

According to the Hasse diagram, the structure matrices of $\vee$ and $\wedge$ can easily be calculated:
\begin{align}\label{2.3.5}
\begin{array}{ccl}
M_{\vee}&=&\d_5[1,1,1,1,1,1,2,1,2,2,1,1,3,3,3,\\
~&~&~~~~1,2,3,4,4,1,2,3,4,5].
\end{array}
\end{align}
\begin{align}\label{2.3.6}
\begin{array}{ccl}
M_{\wedge}&=&\d_5[1,2,3,4,5,2,2,4,4,5,3,4,3,4,5,\\
~&~&~~~~4,4,4,4,5,5,5,5,5,5].
\end{array}
\end{align}

Using the structure matrices of $\vee$ and $\wedge$, we can calculate the structure matrices of the logical functions in (\ref{2.3.2}).
$$
\begin{array}{l}
M_{\vee}ux_2=M_{\vee}({\bf 1}_5^\mathrm{T}\otimes I_5\otimes {\bf 1}_5^\mathrm{T}\otimes I_5)ux\\
:=L_1ux,
\end{array}
$$
where
$$
\begin{array}{ccl}
L_1&=&M_{\vee}({\bf 1}_5^\mathrm{T}\otimes I_5\otimes {\bf 1}_5^\mathrm{T}\otimes I_5)\\
~&=&\d_5[1,1,1,1,1,1,1,1,1,1,1,1,1,1,1,1,1,1,1,1,\\
~&~&\cdots\\
~&~&2,2,2,2,2,3,3,3,3,3,4,4,4,4,4,5,5,5,5,5].\\
\end{array}
$$
$$
\begin{array}{l}
M_{\wedge}x_1x_2=M_{\wedge}({\bf 1}_5^\mathrm{T}\otimes I_5\otimes {\bf 1}_5^\mathrm{T}\otimes I_5)ux\\
:=L_2ux,
\end{array}
$$
where
$$
\begin{array}{ccl}
L_2&=&M_{\wedge}({\bf 1}_5^\mathrm{T}\otimes I_5\otimes {\bf 1}_5^\mathrm{T}\otimes I_5)\\
~&=&\d_5[1,2,3,4,5,1,2,3,4,5,1,2,3,4,5,1,2,3,4,5,\\
~&~&\cdots\\
~&~&5,5,5,5,5,5,5,5,5,5,5,5,5,5,5,5,5,5,5,5].\\
\end{array}
$$
$$
\begin{array}{l}
M_{\wedge}M_{\vee}x_1x_2x_3=M_{\wedge}M_{\vee}({\bf 1}_5^\mathrm{T}\otimes I_{125})ux\\
:=L_3ux,
\end{array}
$$
where
$$
\begin{array}{ccl}
L_3&=&M_{\wedge}M_{\vee}({\bf 1}_5^\mathrm{T}\otimes I_{125})\\
~&=&\d_5[1,2,3,4,5,1,2,3,4,5,1,2,3,4,5,1,2,3,4,5,\\
~&~&\cdots\\
~&~&2,2,4,4,5,3,4,3,4,5,4,4,4,4,5,5,5,5,5,5].\\
\end{array}
$$

It follows that
$$
\begin{array}{ccl}
L&=&L_1*L_2*L_3\\
~&=&\d_{125}[1,2,3,4,5,1,2,3,4,5,1,2,3,4,5,\\
~&~&\cdots,47,47,49,49,50,73,74,73,74,75,\\
~&~&99,99,99,99,100,125,125,125,125,125].\\
\end{array}
$$

Finally, since $J=\{\d_5^4,\d_5^5\}$ is a sub-lattice of $L$ (precisely speaking, $J$ is an ideal of $L$), when all the state variables (i.e., $x_1,x_2,x_3$), take values from $J$, the corresponding states form an invariant subspace. Denote it by $V$. That is,
\begin{align*}
\begin{array}{ccl}
V&=&\left\{x=x_1x_2x_3\;|\; x_1,x_2,x_3\in \{\d_5^4,\d_5^5\}\right\}\\
~&=&\left\{\d_5^4\d_5^4\d_5^4,\d_5^4\d_5^4\d_5^5,\cdots,\d_5^5\d_5^5\d_5^5\right\}\\
~&=&\d_{125}[94,95,99,100,119,120,124,125]\\
~&\in&{\cal L}_{125\times 8}.
\end{array}
\end{align*}

If the control $u=\d_5^4\in J$, we have
$$
M_4=L\d_5^4\in {\cal L}_{125\times 125}.
$$
Then $V$ is $M_4$-invariant. A straightforward computation shows that
$$
M_4V=VE_1,
$$
where
$
E_1=\d_8[1,4,1,4,3,4,4,4].
$

If the control $u=\d_5^5\in J$, we have
$$
M_5=L\d_5^5\in {\cal L}_{125\times 125}.
$$
Then $V$ is $M_5$-invariant. It is also easy to show that
$$
M_5V=VE_2,
$$
where
$
E_2=\d_8[1,4,5,8,3,4,8,8].
$
\end{exa}

\begin{rem}\label{r2.3.6} In both Example \ref{e2.3.4} and Example \ref{e2.3.5} the invariant subspaces come from the ``subspace" structure of the bearing spaces, from which the nodes of networks take values. Such invariant subspaces are independent of the particular networks. 
\end{rem}

\section{Universal Cascading Form}

\subsection{Universal Invariant Subspace}

\begin{dfn}\label{d3.1.1} Assume $B$ is the bearing space of a network with its operators $t_i$, $i=1,2,\cdots,s$. $H\subset B$ and the restrictions of $t_i|_{H}$,  $i=1,2,\cdots,s$ are properly defined, then $H$ is called a bearing subspace of $B$.\footnote{In fact, $H$ is a subuniverse of the universal algebra $B$ \cite{bur81}.}
\end{dfn}

\begin{exa}\label{e3.1.2}
\begin{itemize}
\item[(i)] Consider a ring $\Z_6$, assume $H=\{0,3\}$ which is a sub-ring of $\Z_6$. Then the operators $\oplus|_{H}$ and $\odot|_{H}$ are properly defined. So, if $\Z_6$ is a bearing space of a network, then $H$ is a bearing subspace of the network.
\item[(ii)] Consider a lattice $L$ with its Hasse diagram as in Fig. \ref{Fig.1.2.1}. If $L$ is a bearing space of a network,  Then $H_0=\{D\}$, $H_1=\{B,D\}$, $H_2=\{A,B,D\}$, as sub-lattices of $L$, are all bearing subspaces.
\end{itemize}
\end{exa}

Consider $k$-valued network (\ref{2.1.1}) (or $k$-valued control network (\ref{2.1.2})). Assume its bearing space is $B$ and $H\subset B$ is
a bearing subspace. The state subspace corresponding to $H$ is defined by
$$
{\cal X}|_{H}=\{(X_1,X_2,\cdots,X_n)\in {\cal X}\;|\; X_i\in H,\;i\in [1,n]\}.
$$

The following proposition comes from the definition of bearing space.

\begin{prp}\label{p3.1.3} Assume $k$-valued network (\ref{2.1.1}) (or $k$-valued control network (\ref{2.1.2})) has its bearing space $B$, and $H\subset B$ is a bearing subspace. Then the ${\cal X}|_{H}$ is the state invariant subspace of (\ref{2.1.1}).
It is also a control-invariant subspace of (\ref{2.1.2})). Moreover,  ${\cal X}|_{H}$ is independent of a particular network. It is, therefore, called the universal state invariant subspace.
\end{prp}

\begin{rem}\label{r3.1.4}
\begin{itemize}
\item[(i)] Assume
$$
H=\left\{\d_k^{r_1},\d_k^{r_2},\cdots,\d_k^{r_s}\right\}.
$$
Then
\begin{align}\label{3.1.1}
{\cal X}|_{H}=\left\{\ltimes_{i=1}^n\d_k^{j_i}\;|\; j_i\subset\{r_1,r_2,\cdots,r_s\},\;i\in [1,n]\right\}.
\end{align}
\item[(ii)] To see ${\cal X}|_{H}$ is also a control-invariant state subspace, we simply set $u_j\in H$, $j\in [1,m]$, the conclusion follows.
\end{itemize}
\end{rem}

\subsection{Cascading Normal Form}

\begin{prp}\label{p3.2.1} Let $H_1\subset H_2\subset \cdots\subset H_s=B$ be a set of nested bearing subspaces.
\begin{itemize}
\item[(i)] Consider $k$-valued network (\ref{2.1.1}) with $B$ as its bearing space. Then there exists a coordinate transformation $Z=T(X)$, such that under the coordinate frame $Z$, the ASSR of (\ref{2.1.1}) becomes
\begin{align}\label{3.2.1}
z(t+1)=Mz(t),
\end{align}
where $M$ has the cascading form (upper block triangular form) as
\begin{align}\label{3.2.2}
M=\begin{bmatrix}
D_{1,1}&D_{1,2}&\cdots&D_{1,s}\\
0&D_{2,2}&\cdots&D_{2,s}\\
\vdots&~&~&~\\
0&0&\cdots&D_{ss}\\
\end{bmatrix},
\end{align}
where the nested diagonal square blocks
$$
D_{1,1},~\begin{bmatrix}D_{1,1}&D_{1,2}\\0&D_{2,2}\end{bmatrix},~\cdots,~M
$$
correspond to the nested $H_1\subset H_2\subset \cdots \subset H_s$ respectively.

\item[(ii)]   Consider $k$-valued control network (\ref{2.1.2}) with $B$ as its bearing space, and assume $u\in H_r$. Then there exists a coordinate transformation $Z=T(X)$, such that under the coordinate frame $Z$, the ASSR of (\ref{2.1.2}) becomes
\begin{align}\label{3.2.3}
z(t+1)=\tilde{L}(u)z(t),
\end{align}
\begin{align}\label{3.2.4}
\tilde{L}(u)=\begin{bmatrix}
L_{1,1}&L_{1,2}\\
0&L_{2,2}\\
\end{bmatrix},
\end{align}
where $L_{1,1}$ corresponds to $H_r$.
\end{itemize}
\end{prp}

{\it Proof:}
\begin{itemize}
\item[(i)] We give constructive proof.
Define
$$
T_i:=\left\{\ltimes_{j=1}^n x^i_j\;|\;x^i_j\in H_i \right\}.
$$

It follows from the construction  that if $X\in T_i$, $Y\in T_j\T_i$, $j>i$, then
\begin{align}\label{3.2.5}
\left<X,Y\right>=0.
\end{align}

Then we construct a $k^n\times k^n$ matrix as
\begin{align}\label{3.2.6}
T:=\left[T_1\;|\;T_2\backslash T_1\;|\;T_3\backslash T_2\;|\;\cdots\;|\;T_s\backslash T_{s-1}\right].
\end{align}

Using some properties of the STP, a straightforward computation verifies that
 $T$ is a coordinate transformation, and
$$
T^{-1}=T^\mathrm{T}.
$$
Moreover, set $x=Tz$, then it is easy to verify that
\begin{itemize}
    \item $T_i$ and $T_j\backslash T_i$ are orthogonal, where $j>i$.
    \item $MT_i\subset T_i$.
\end{itemize}
These two facts lead to the conclusion that
under the coordinate frame $z$ the transition matrix $T^\mathrm{T}MT$ has the form (\ref{3.2.2}).

\item[(ii)] Since $u\in H_r$, $H_r$ is $L(u)$ invariant. The conclusion becomes a special case of (i).
\end{itemize}
\hfill $\Box$

\begin{exa}\label{e3.2.2} Consider a $4$-valued  network
\begin{align}\label{3.2.7}
\begin{cases}
X_1(t+1)=X_1(t)\wedge X_2(t),\\
X_2(t+1)=X_1(t)\vee X_2(t).
\end{cases}
\end{align}
Assume its bearing space is a lattice $L$ with its Hasse diagram
as in Fig. \ref{Fig.1.2.1}. Denote by $A=\d_4^1$, $B=\d_4^2$, $C=\d_4^3$, and $D=\d_4^4$.
Skipping the tedious calculation, which is similar to Example \ref{e2.3.5},  the ASSR form of (\ref{3.2.6}) can be obtained as
\begin{align}\label{3.2.8}
x(t+1)=Mx(t),
\end{align}
where
$$
M=\d_{16}[1,5,9,13,5,6,13,14,9,13,11,15,13,14,15,16].
$$

It is easy to calculate that
$$
\begin{array}{ccl}
T_1&=&\{\d_4^4\d_4^4\},\\
T_2\backslash T_1&=&\{\d_4^4\d_4^2, \d_4^2\d_4^4, \d_4^2\d_4^2\},\\
T_3\backslash T_2&=&\{\d_4^4\d_4^1, \d_4^1\d_4^4, \d_4^2\d_4^1, \d_4^1\d_4^2\},\\
T_4\backslash T_3&=&\{\d_4^1\d_4^1, \d_4^4\d_4^3, \d_4^3\d_4^4, \d_4^2\d_4^3,\\
~&~& \d_4^3\d_4^2, \d_4^1\d_4^3, \d_4^3\d_4^1, \d_4^3\d_4^3\}.\\
\end{array}
$$
Then we got
$$
\begin{array}{ccl}
T&=&\left[T_1,T_2\backslash T_1,T_3\backslash T_2,T_4\backslash T_3 \right]\\
~&=&\d_{16}[16,14,8,6,13,4,5,2,1,15,12,7,10,3,9,11];
\end{array}
$$
Finally, we set $x=Tz$ and calculate the transition matrix under the $z$ coordinate frame as
\begin{align}\label{3.2.9}
z(t+1)=T^\mathrm{T}MTz(t):=\tilde{M}z(t),
\end{align}
where
\begin{align}\label{3.2.901}
\tilde{M}=\d_{16}[1,2,2,4,5,5,7,7,9,10,10,5,5,15,15,16],
\end{align}
which is in cascading form with
$$
\begin{array}{lll}
[1] &\rightarrow & H_1,\\
\d_{4}[1,2,2,4]&\rightarrow& H_2,\\
\d_8[1,2,2,4,5,5,7,7]&\rightarrow& H_3,\\
\tilde{M}&\rightarrow& B.\\
\end{array}
$$
\end{exa}

\begin{rem}\label{r3.2.3} Precisely speaking, bearing subspaces depend also on network dynamics in certain sense. For instance, in Example \ref{e3.2.2}, only $\wedge$ and $\vee$ are used in dynamic equation (\ref{3.2.7}), hence all the sun-lattices of the lattice  $L$ in Fig. \ref{Fig.1.2.1} are its bearing subspaces.
\begin{itemize}
\item[(i)] Since $L$ is a complementation lattice, each element $a$ has its complement $\neg a$ \footnote{A lattice is called a complementation lattice if $L$ has a maximum element $\J$ and minimum element $\0$ such that for each element $a\in L$ there is an element, called the complement of $a$ and denoted by $\neg a$, such that
    $$
    a\vee \neg a=\J,\quad a\wedge \neg a=\0.
    $$
    }. 
    Now if $\neg$ is used in the dynamic equation (\ref{3.2.7}), then only the complementation sub-lattice can be considered as a bearing subspace. That is, we have only $H_1=\{A, D\} \subset L$ as its bearing subspace.
\item[(ii)] If we consider conjunctive networks over $L$, which are similar to conjunctive Boolean networks \cite{gao18},
 since $(L,\wedge)$ is a semi-lattice, any sub-semi-lattices can be considered as its bearing subspace. For instance, a nested bearing subspaces can be $H_1=\{D\}$, $H_2=\{B,D\}$ $H_3=\{D,B,C\}$, $H_4=L$.
 \end{itemize}
 \end{rem}

\section{Duality of Invariant Subspaces}

\subsection{Invariant Subspace vs Dual Invariant Subspace}

Assuming the bearing space $B$ of a finite-valued network has a ``zero", then we consider the duality of invariant subspaces with invariant dual subspaces. We give a precise definition.

\begin{dfn}\label{d4.1.1}Given a bearing space $B$ of a finite-valued network $\Sigma$, as described by (\ref{2.1.1}) with its ASSR (\ref{2.1.4}). Assume there is a $\theta\in B$, and a binary operator $\odot$ over $B$, such that
$$
b\odot \theta=\theta\odot b=\theta,\quad \forall b\in B,
$$
then $B$ is said to have a zero element $\theta$, and the operator $\odot$ is considered as a product over $B$.
\end{dfn}

\begin{exa}\label{e4.1.2}
\begin{itemize}
\item[(i)] Assume $B$ is a finite ring (including finite field), then $B$ is with zero, where $\theta=0$ and $\odot$ is the product of the ring.
\item[(ii)] Assume $B$ is a $k$-valued logic (including Boolean logic), then $B$ is with zero, where $\theta=0$ with product $\wedge$. (If you wish, you can also set $\theta=1$ with product $\vee$.) 
\item[(iii)] Assume $B$ is a finite lattice, then $B$ is with zero, where $\theta=\min$ (smallest element) with product $\inf$.
\end{itemize}
\end{exa}

\begin{dfn}\label{d4.1.3} Given a bearing space $B$ of a finite-valued network $\Sigma$, as described by (\ref{2.1.1}), and assume $B$ has zero $\theta$.
\begin{itemize}
\item[(i)] Let ${\cal G}^*={\cal F}_{\ell}(g_1,g_2,\cdots,g_r)$ be a dual subspace of ${\cal X}^*$. Then
\begin{align}\label{4.1.1}
{{\cal G}^*}^{\perp}:=\{X\in {\cal X}\;|\;g_i(X)=\theta,\; i\in [1,r]\}
\end{align}
is called the state space perpendicular to ${\cal G}^*$.

\item[(ii)] Let ${\cal V}=\{X_1,X_2,\cdots,X_s\}$ be a subspace of ${\cal X}$. Then
\begin{align}\label{4.1.2}
{\cal V}^{\perp}:=\{f\in {\cal X}^*\;|\;f(X_i)=\theta,\; i\in [1,s]\}
\end{align}
is called the dual state space perpendicular to ${\cal V}$.
\end{itemize}
\end{dfn}

The following result shows the relationship between invariant subspace and invariant dual subspace.

\begin{thm}\label{t4.1.4} Consider $k$-valued network (\ref{2.1.1}) with its ASSR (\ref{2.1.4}) and assume its bearing space $B$ has zero element $\theta$.
\begin{itemize}
\item[(i)] Assume ${\cal G}^*$ is an  $M$-invariant dual subspace. Then its perpendicular subspace ${\cal V}={{\cal G}^*}^{\perp}$ is an
$M$-invariant subspace.
\item[(ii)] Assume ${\cal V}$ is an  $M$-invariant subspace. Then its perpendicular dual subspace ${\cal G}^*={\cal V}^{\perp}$ is an
$M$-invariant dual subspace.
\end{itemize}
\end{thm}

{\it Proof:}
We prove (i) only,  the proof of (ii) is similar.

Let $x\in {{\cal G}^*}^{\perp}$. Consider $Mx$: Since $G\in {\cal L}_{k^r\times k^n}$ is $M$-invariant, there exists $H$ such that
$$
GMx=HGx.
$$
Now since $HG\subset \Span\{Row(G)\}$, $HGx=\theta^r$. Hence $Mx\in  {{\cal G}^*}^{\perp}$.
\hfill $\Box$

The following is an immediate consequence.

\begin{cor}\label{c4.1.5} Given a bearing space $B$ of a finite control network $\Sigma$, as described by the $k$-valued control network (\ref{2.1.2}) with its ASSR (\ref{2.1.4}) and assume its bearing space $B$ has zero element $\theta$.
\begin{itemize}
\item[(i)] Assume ${\cal G}^*$ is a control-invariant dual subspace. Then its perpendicular subspace ${\cal V}={{\cal G}^*}^{\perp}$ is a control
invariant subspace.
\item[(ii)]  Assume ${\cal V}$ is a control-invariant subspace. Then its perpendicular dual subspace ${\cal G}^*={\cal V}^{\perp}$ is a
control-invariant dual subspace.
\end{itemize}
\end{cor}

We use a simple example to verify Theorem \ref{t4.1.4}.

\begin{exa}\label{e4.1.6} Consider a Boolean network
\begin{align}\label{4.1.3}
\begin{cases}
X_1(t+1)=X_1(t)\vee X_2(t),\\
X_2(t+1)=X_1(t)\wedge X_2(t),\\
X_3(t+1)=X_1(t)\lra X_3(t).
\end{cases}
\end{align}
Then it is obvious that ${\cal G}^*={\cal F}_{\ell}(X_1, X_2)$ is a node invariant subspace. The ASSR of (\ref{4.1.3}) can easily be calculated as
\begin{align} \label{4.1.4}
x(t+1)=Mx(t),
\end{align}
where
$
M=\d_8[1,2,3,4,4,3,8,7].
$

Let the structure matrix of ${\cal G}^*$ be $G$. Then it is easy to calculate that
$$
G=\d_4[1,1,2,2,2,2,4,4].
$$
A straightforward verification shows that
$$
GM=HG,
$$
where $H=I_4$.

Consider $E:={{\cal G}^*}^{\perp}$. Then
$$
\begin{array}{ccl}
E&=&\{\d_2^2\d_2^2x_3\;|\;x_3\in \D_2\}\\
~&=&\{\d_8^7,\d_8^8\}.
\end{array}
$$
It is ready to verify that $$ME=E\d_2[2,1]\subset E.$$

\end{exa}

\subsection{Universal Dual Invariant Subspace}

Using the duality of invariant subspaces with dual invariant subspaces, one sees easily that the universal invariant subspaces also have their corresponding (perpendicular) dual invariant subspaces. According to Theorem \ref{t4.1.4}, it is obvious that the dual invariant subspaces are also independent of the particular network dynamics. Hence we have the following result.

\begin{prp}\label{p2.2.1} Consider the $k$-valued network (\ref{2.1.1}). Assume its bearing space $B$ has zero $\theta$. If ${\cal V}$ is a universal invariant subspace, then ${\cal G}^*={\cal V}^{\perp}$ is also universal. That is, it is independent of particular network dynamics.
\end{prp}

\begin{exa}\label{e2.2.2} Recall Example \ref{e3.2.2} and observe Fig.\ref{Fig.1.2.1}. There is a sequence of universal invariant subspaces:
$$
H_1\subset H_2\subset H_3\subset H_4=L,
$$
where $H_i$ corresponds to $T_i,~i=1,2,3,4$.

Then we have a sequence of universal dual invariant subspaces as
$$
{\cal G}_4^*\subset {\cal G}_3^*\subset {\cal G}_2^*\subset {\cal G}_1^*,
$$
where $
{\cal G}_i^*=H_i^{\perp},~i=1,2,3,4.
$

Then it is clear that the elements of ${\cal G}_i^*$, denoted their vector forms by $g_i$, $i=1,2,3,4$, are
$$
\begin{array}{ccl}
g_4&=&\d_4[4,4,4,4,4,4,4,4,4,4,4,4,4,4,4,4],\\
g_3&=&\d_4[4,4,*,4,4,4,*,4,*,*,*,*,4,4,*,4],\\
g_2&=&\d_4[*,*,*,*,*,4,*,4,*,*,*,*,*,4,*,4],\\
g_1&=&\d_4[*,*,*,*,*,*,*,*,*,*,*,*,*,*,*,4],\\
\end{array}
$$
where $*\in\{1,2,3,4\}$ stands for arbitrary value.

Using the coordinate transformation
$$
\begin{array}{ccl}
T&=&\d_{16}[16,14,8,6,13,4,5,2,1,15,12,7,10,3,9,11],\\
\tilde{g}_4&:=&g_4T=\d_4[4,4,4,4,4,4,4,4,4,4,4,4,4,4,4,4],\\
\tilde{g}_3&:=&g_3T=\d_4[4,4,4,4,4,4,4,4,4,*,*,*,*,*,*,*],\\
\tilde{g}_2&:=&g_2T=\d_4[4,4,4,4,*,*,*,*,*,*,*,*,*,*,*,*],\\
\tilde{g}_1&:=&g_1T=\d_4[4,*,*,*,*,*,*,*,*,*,*,*,*,*,*,*].\\
\end{array}
$$
It is ready to verify that ${\cal G}_i^*=\{\tilde{g}_i\}~i=1,2,3,4$ are $\tilde{M}$-invariant, where $\tilde{M}$ is given in (\ref{3.2.901}).
\end{exa}

\section{Conclusion}

Though the concept of ``invariant subspace" has been used in the STP-based study of Boolean (control) networks for a considerable long time, it has not been clearly defined and there are some unclear or murky issues. In this paper, the two kinds of (control) invariant subspaces of $k$-valued networks have been clearly defined. They are the state invariant and dual invariant subspaces of the state space and the dual space respectively. Then their properties have been investigated.

First, for invariant dual subspaces, two algorithms have been proposed to verify if a dual subspace of a network is an invariant subspace and a dual subspace of a control network is a dual control invariant subspace respectively. Second, for state invariant subspaces, the relationship between the topological structure and the state invariant subspaces is revealed. Then a special kind of state invariant subspaces, called universal subspaces, are considered. They are caused by the structure of bearing space and are independent of the dynamics of individual networks. Based on universal invariant subspaces the universal normal form is revealed. Finally, the duality of state invariant subspaces and dual invariant subspaces is obtained, which shows that if a dual subspace is invariant, then its perpendicular state subspace is also invariant, and vice versa. 

Since invariant subspace is a fundamental concept and a useful tool in many control problems of networks, this paper may help to build a solid theoretical foundation for further investigation and applications.

\appendix
\section{Semi-tensor Product of Matrices}
The semi-tensor product of matrices is defined as follows \cite{che11,che12}:

\begin{dfn}\label{da1.1.1}  Let $M\in {\cal M}_{m\times n}$, $N\in {\cal M}_{p\times q}$, and $t=\lcm\{n,p\}$ be the least common multiple of $n$ and $p$.
The semi-tensor product (STP) of $M$ and $N$ is defined as
\begin{align}\label{a1.1.1}
M\ltimes N:= \left(M\otimes I_{t/n}\right)\left(N\otimes I_{t/p}\right)\in {\cal M}_{mt/n\times qt/p},
\end{align}
where $\otimes$ is the Kronecker product.
\end{dfn}

 The STP of matrices is a generalization of the conventional matrix product, and all the computational properties of the conventional matrix product remain available. Throughout this paper, the default matrix product is STP, so the product of two arbitrary matrices is well defined, and the symbol $\ltimes$ is mostly omitted.

The following properties will be used in the sequel:

\begin{prp} \label{pa1.1.2} Let $X\in \mathbb{R}^t$. Then for a given matrix $A$ we have
\begin{align}\label{a1.1.2}
XA=(I_t\otimes A)X.
\end{align}
\end{prp}

A swap matrix $W_{[m,n]}\in {\cal M}_{mn\times mn}$ is defined as

\begin{align}\label{a1.1.3}
W_{[m,n]}:=[I_n\otimes \d_m^1,I_n\otimes \d_m^2,\cdots,I_n\otimes \d_m^m].
\end{align}

It is used to swap two vectors.

\begin{prp} \label{pa1.1.3} Let $X\in \mathbb{R}^m$, $Y\in \mathbb{R}^n$ be two column vectors. Then
\begin{align}\label{a1.1.4}
W_{[m,n]}XY=YX.
\end{align}
\end{prp}

A power reducing matrix ${PR}_{k}\in {\cal M}_{k^2\times k}$ is defined as
\begin{align}\label{a1.1.5}
{PR}_{k}:=\diag(\d_k^1,\d_k^2,\cdots,\d_k^k).
\end{align}

It is used to reduce the power of a vector.

\begin{prp} \label{pa1.1.4} Let $X\in \mathbb{R}^k$. Then
\begin{align}\label{a1.1.6}
{PR}_kX=X^2.
\end{align}
\end{prp}
Using STP to $k$-valued logical mappings, we have the following result.
\begin{prp} \label{pa1.1.5}
Let $f:\D_k^n\rightarrow \D_k$. Then there exists a unique matrix, called the structure matrix of $f$ and denoted by $M_f\in {\cal L}_{k\times k^n}$, such that
\begin{align}\label{a1.1.7}
f(x_1,x_2,\cdots,x_n)=M_f\ltimes_{i=1}^nx_i,\quad x_i\in \D_k,\; i\in [1,n].
\end{align}
\end{prp}

\begin{dfn} \label{da1.1.6} Let $A\in {\cal M}_{p\times n}$ and $B\in {\cal M}_{q\times n}$. The Khatri-Rao product of $A$ and $B$, denoted by $A*B$, is defined as follows.
\begin{align}\label{a1.1.8}
\begin{array}{ccl}
A*B&=&\left[\Col_1(A)\Col_1(B), \Col_2(A)\Col_2(B),\cdots,\right.\\
~&~&\left.\Col_n(A)\Col_n(B)\right]\in {\cal M}_{pq\times n}.
\end{array}
\end{align}
\end{dfn}

\section{Some Bearing Spaces}

In this section, we give a brief introduction to some basic algebraic structures, which are used to describe structures of some commonly used bearing spaces of $k$-valued networks.

\begin{dfn}\label{da1.2.1} \cite{hun74}  A set $G\neq \emptyset$ with a binary operator $*:G\times G$ is a group, if
\begin{itemize}
\item[(i)]
$
a*(b*c)=(a*b)*c,\quad a,b,c\in G;
$
\item[(ii)] there exists an identity $e\in G$, such that
$$
a*e=e*a=a,\quad \forall a\in G;
$$
\item[(iii)] for each $a\in G$ there exists a inverse $a^{-1}$, such that
$$
a*a^{-1}=a^{-1}*a=e.
$$
\end{itemize}
If only (i) is satisfied, $G$ is a semi-group. If only (i) and (ii) are satisfied, $G$ is a monoid (or semi-group with identity).

In addition, if a group satisfies
$$
a*b=b*a,\quad a,b\in G,
$$
then $G$ is called an Abelian group.
\end{dfn}

\begin{dfn}\label{da1.2.2} \cite{hun74} A set $R\neq \emptyset$ with two binary operators $+$ and $*$ is called a ring, denoted by $(R,+,*)$,  if
\begin{itemize}
\item[(i)] $(R,+)$ is an Abelian group with identity $\0$,
\item[(ii)] $(R,*)$ is a semi-group,
\item[(iii)] the following distributive rule is satisfied:
$$
\begin{array}{l}
(a+b)*c=a*c+b*c,\\
c*(a+b)=c*a+c*b,\quad a,b,c\in R.
\end{array}
$$
\end{itemize}
\end{dfn}

When $(R,*)$ is with identity ${\bf 1}$, it is called a ring with identity. If $a*b=b*a$ for $a,b\in R$, $R$ is called a commutative ring. If $E\subset R$ and $(E,+,*)$ is also a ring, $E$ is said to be a sub-ring of $R$. If $E$ is a sub-ring of $R$ and for any $r\in \R$ and $e\in E$
$$
r*e, ~e*r\in E,
$$
then $E$ is called an ideal of $R$.

\begin{dfn}\label{da1.2.3} \cite{hun74} A set $(F,+,*)$ is a field, if it is a ring, and
 $(F\backslash\{\0\},*)$ is also an Abelion group.
\end{dfn}

\begin{dfn}\label{da1.2.4} \cite{bur81} A partial ordered set $(L, \prec)$ is called a lattice, if for any two elements
$a,b\in L$, there exists a least upper bound, denoted by $\sup(a,b)=a\vee b$, and a greatest lower bound, denoted by $\inf(a,b)=a\wedge b$.
\end{dfn}

If $(L,\prec)$ is a lattice, $H\subset L$ and $(H,\prec)$ is also a lattice, then $H$ is said to be a sub-lattice of $L$. A sub-lattice $H$ is called an ideal, if for any $a\in L$ there exists an $h\in H$ such that $a\prec h$ implies $a\in H$.

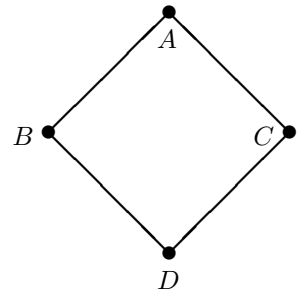
\begin{figure}
\centering
\setlength{\unitlength}{8mm}
\begin{picture}(5,5)(-1,-1)\thicklines
\put(0,2){\circle*{0.2}}
\put(2,0){\circle*{0.2}}
\put(2,4){\circle*{0.2}}
\put(4,2){\circle*{0.2}}
\put(0,2){\line(1,1){2}}
\put(0,2){\line(1,-1){2}}
\put(4,2){\line(-1,1){2}}
\put(4,2){\line(-1,-1){2}}
\put(1.8,3.4){$A$}
\put(-0.6,1.8){$B$}
\put(1.8,-0.6){$D$}
\put(3.4,1.8){$C$}
\end{picture}
\caption{Hasse Diagram of $L$ \label{Figa.1.2.1}}
\end{figure}

A lattice can be described by a figure called the Hasse diagram. For instance, Fig. \ref{Figa.1.2.1} shows a lattice $L$, where
$L=\{A,B,C,D\}$. It is easy to verify that $A\vee B=A$, $A\wedge B=B$, $B\vee C=A$, $B\wedge C=D$, etc.

\vskip 5mm

\begin{thebibliography}{00}
%
\bibitem{ada03} A. Adamatzky, On dynamically non-trivial three-valued logics: oscillatory and bifurcatory spaces, {\it Chaos Solitons Fractals}, Vol. 18, 917-936, 2003.
%
\bibitem{aku07} T. Akutsi, M. Hayashida, W. Ching, M. Ng, Control of Boolean networks: hardness results and algorithms for three structured networks, {\it J. Theor. Biol.}, Vol. 244, No. 4, 670-769, 2007.
%
\bibitem{bur81} S. Burris, H.H. Sankappanavar, {\it A Course in Universal Algebra}, Springer-verlag, New York, 1981.
%
\bibitem{bry84} R.E. Bryant, A switch-level model and simulator for MOS digital systems, {\it IEEE Trans. Comput.}, Vol. C-33, No. 2, 160-177, 1984.
%
\bibitem{che10} D. Cheng, H. Qi, State space analysis of Boolean networks, {\it IEEE Trans. Neural Netw.}, Vol. 21, No .4, 584-594, 2010.
%
\bibitem{che11} D. Cheng, H. Qi, Z. Li, {\it Analysis and Control of Boolean Networks: A Semi-tensor Product Approach}, Springer, London, 2011.
%
\bibitem{che11b} D. Cheng, Disturbance decoupling of Boolean control networks, {\it IEEE Trans. Aut. Contr.}, Vol. 56, No. 1, 2-10, 2011.
%
\bibitem{che12} D. Cheng, H. Qi, Y. Zhao, {\it An Introduction to Semi-tensor Product of Matrices and Its
Applications}, World Scientific, Singapore, 2012.
%
\bibitem{che15} D. Cheng, F. He, H. Qi, T. Xu, Modeling analysis and control of networked evolutionary games, {\it IEEE Trans. Aut. Contr.}, Vol. 60, No. 9, 2402-2415, 2015.
%
\bibitem{che21} D. Cheng, Y. Wu, G. Zhao, S. Fu, A comprehensive survey on STP approach to finite games, {\it J. Sys. Sci. Compl.}, Vol. 34, No. 5, 1666-1680, 2021.
%
\bibitem{che22} D. Cheng, L. Zhang, D. Bi, Invariant subspace approach to Boolean (control) networks, {\it IEEE Trans. Aut. Contr.}, (on line) ieeexplore.ieee.org/document/9775567, 2022.
%
\bibitem{che22b} D. Cheng, Z. Ji, On networks over finite rings,  {\it J. Franklin Inst.}, Vol. 359, No. 14, 7562-7599, 2022.
%
\bibitem{dat03} A. Datta, A. Choudhary, M.L. Bittner, E.R. Dougherty, External control in Markovian genetic regulatory networks, {\it Mach. Learn.}, Vol. 52, 169-191, 2003.
%
\bibitem{eps60} G. Epstein, The lattice theory of Post algebras, {\it Trans. Am. Soc.}, Vol. 95, 300-317, 1960.
%
\bibitem{for13} E. Fornasini, M.E. Valcher, Observability, reconstructibility and state observers of Boolean control networks, {\it IEEE Trans. Aut. Contr.}, Vol. 58, No. 6, 1390-1401, 2013.
%
\bibitem{gao18} Z. Gao, X. Chen, T. Ba\c{s}ar, Stability structures of conjunctive Boolean networks, {\it Automatica}, Vol.  89, 8-20, 2018.
%
\bibitem{guo13} P. Guo, Y. Wang, H. Li, Algebraic formulation and strategy optimization for a class of evolutionary networked games via semi-tensor product method, {\it Automatica}, Vol. 49, 3384-3389, 2013.
%
\bibitem{guo15} Y. Guo, P. Wang, W. Gui, C. Yang, Set stability and set stabilization of Boolean control networks based on invariant subsets, {\it Automatica}, Vol. 61, 106,112, 2015.
%
\bibitem{guo17} Y. Guo, Y. Ding, D. Xie, Invariant subset and set stability of Boolean networks under arbitrary switching signals, {\it IEEE Trans. Aut. Contr.}, Vol. 62, No. 8, 4209-4214, 2017.
%
\bibitem{guo19} Y. Guo, R. Zhou, Y. Wu, W. Gui, C. Yang, Stability and set stability in distribution of probabilistic Boolean networks, {\it IEEE Trans. Aut. Contr.}, Vol. 64, No. 2, 736-742, 2019.
%

\bibitem{hun74} T.W. Hungerford, {\it Algebra}, Springer-Verlag, Now York, 1974.
%
\bibitem{ide01} T. Ideker, T. Galitski, L. Hood, A new approach to decoding life: systems biology, {\it Annu. Rev. Genom. Hum. Genet}, Vol. 2, 343-372, 2001.
%
\bibitem{isi95} A. Isidori, {\it Non-linear Control Systems}, 3rd ed., Springer-Verlag, New York, 1995.
%
\bibitem{jipr} Z. Ji, D. Cheng, Control networks over finite lattices, preprint, arxiv:2208.03716, 2022.
%
\bibitem{jon12} J. de Jong, Modeling and simulation of genetic regulatory systems: a literature review, {\it J. Comput. Biol.}, Vol. 9, No. 1, 67-103, 2012.
%
\bibitem{kal60} R.E. Kalman, On the general theory of control systems, in Automatic and Remote control (Proc. 1st Int. Congr. Internat. Fed. Aut. Contr., Moscow, 1960), Butterworths,  London, Vol. 1, 481-492, 1961.
%
\bibitem{kau69} S.A. Kauffman, Metabolic stability and epigenesis in randomly constructed genetic nets, {\it J. Theoret. Biol.}, Vol. 22, 437-467, 1969.
%
\bibitem{li10} Z. Li, D. Cheng, Algebraic approach to dynamics of multi-valued networks, {\it Int. J. Bifurc. Chaos}, Vol. 20, No. 3, 561-582, 2010.
%
\bibitem{li16} X. Li, M. Chen, H. Su, C. Li, Consensus networks with switching topology and time-delays over finite fields, Automatica 68 (2016) 39-43.
%
\bibitem{li18} H. Li, G. Zhao, P. Guo, {Analysis and Control of Finite-Valued Systems}, CRC Press, 2018.
%
\bibitem{li19} Y. Li, H. Li, X. Ding, G. Zhao, Leader-follower consensus of multiagent systems with time delays over finite fields, {IEEE Trans. Cyb.} 49 (8) (2019) 3203-3208.
%
\bibitem{liu19} Z. Liu, D. Cheng, Canonical form of Boolean networks, {\it Proc. 38th CCC}, 1801-1806, 2019.
%
\bibitem{liu20} A. Liu, H. Li, On feedback invariant subspace of Boolean control networks, {\it Sci. China Inf. Sci.}, 63, 229201, 2020.
%
\bibitem{men20} M. Meng, X. Li, G. Xiao, Synchronization of networks over finite fields, {Automatica} 115 (2020) 108877.
%
\bibitem{pas14} F. Pasqualetri, D. Borra, F. Bullo, Consensus networks over finite fields, {Automatica} 50 (2) (2014) 349-358.
%
\bibitem{pos21} E.L. Post, Introduction to a general theory of elementary propositions, {\it AM.J. Math.}, Vol. 43, 163-185, 1921.
%
\bibitem{rem06} E. Remy, P. Ruet, D. Thieffry, Positive or negative regulatory circuit inference from multilevel dynamics, in {\it Positive Systems: Theory and Applications}, Lecture Notes in Control and Information Science, Vol. 341, Springer, Berlin, 263-270, 2006.
%
\bibitem{ric07} A. Richard, J.P. Comet, Necessary conditions for multistationarity in discrete dynamical systems, {\it Disc. Appl. Math.}, Vol. 155, 2403-2413, 2007.
\bibitem{ros03} K.A. Ross, C.R.B. Wright, {\it Discrete Mathematics}, 5th Ed., Prentice Hall, New Jersey, 2003.
%
\bibitem{sun12} S. Sundaram, C.N. Hadjicostis, Structural controllability and observability of linear systems over finite fields with applications to multi-agent systems, {IEEE Trans. Aut. Contr.} 58 (1) (2012) 60-73.
%
\bibitem{tho01} R. Thomas, M. Kaufman, Multistationrity, the basis of cell differentiation and memory, I\& II, {\it Chaos}, Vol. 11, 170-195, 2001.
%
\bibitem{won70} W.M. Wonham, {\it Linear Multivariable Control, A Geometric Approach}, Springer-Verlag, Berlin, 1974.
%
\bibitem{zha18} X. Zhang, Y. Wang, D. Cheng, Incomplete logical control system and its application to some intellectual problems, {\it Asian J. Contr.}, Vol. 20, No. 2, 697-706, 2018.
%
\bibitem{zha20} X. Zhang, Y. Wang, D. Cheng, Output tracking of Boolean control networks, {\it IEEE Trans. Aut. Contr.}, Vol. 65, No. 6, 2730-2735, 2020.
%
\bibitem{zhu16} J. Zhu, P. L$\ddot{u}$, Regular subspaces and invariant subspaces, {\it IET Contr. Theory Appl.}, Vol. 10, No. 5, 504-508, 2016.
%
\bibitem{zou15} Y. Zou, J. Zhu, Kalman decomposition for Boolean control networks, {\it Automatica}, Vol. 54, 65-71, 2015.
%
\bibitem{zho19} R. Zhou, Y. Guo, Set reachability and observability of probabilistic Boolean networks, {\it Automatica}, Vol. 106, 230-241, 2019.
\end{thebibliography}
\end{document}